\documentclass{article}

\usepackage{PRIMEarxiv}

\usepackage[utf8]{inputenc} 
\usepackage[T1]{fontenc}    
\usepackage{hyperref}       
\usepackage{url}            
\usepackage{booktabs}       
\usepackage{amsfonts}       
\usepackage{nicefrac}       
\usepackage{microtype}      
\usepackage{lipsum}
\usepackage{fancyhdr}       
\usepackage{graphicx}       
\graphicspath{{media/}}     
\usepackage{hyperref}
\usepackage[table,xcdraw]{xcolor}

\title{The Effects of Data Split Strategies on the Offline Experiments for CTR Prediction
}

\author{
  Ramazan Tarik Turksoy \\
  Huawei Turkey R\&D Center \\
  Istanbul, Turkey\\
  \texttt{\ ramazan.tarik.turksoy1@huawei.com} \\
   \And
  Beyza Turkmen \\
  Huawei Turkey R\&D Center \\
  Istanbul, Turkey\\
  \texttt{\ beyza.turkmen2@huawei-partners.com} \\
}

\begin{document}
\maketitle

\begin{abstract}
Click-through rate (CTR) prediction is a crucial task in online advertising to recommend products that users are likely to be interested in. To identify the best-performing models, rigorous model evaluation is necessary. Offline experimentation plays a significant role in selecting models for live user-item interactions, despite the value of online experimentation like A/B testing, which has its own limitations and risks. Often, the correlation between offline performance metrics and actual online model performance is inadequate. One main reason for this discrepancy is the common practice of using random splits to create training, validation, and test datasets in CTR prediction. In contrast, real-world CTR prediction follows a temporal order. Therefore, the methodology used in offline evaluation, particularly the data splitting strategy, is crucial. This study aims to address the inconsistency between current offline evaluation methods and real-world use cases, by focusing on data splitting strategies. To examine the impact of different data split strategies on offline performance, we conduct extensive experiments using both random and temporal splits on a large open benchmark dataset, Criteo.
\end{abstract}

\section{Introduction}

The digital landscape thrives on connecting companies and customers. Companies leverage advertising to reach their target audience while customers rely on recommendation systems to navigate the online marketplace, efficiently discovering products and services that align with their needs. However, a crucial challenge remains: improving Click-through Rate (CTR) prediction within recommendation systems. This study addresses this challenge by proposing a solution to improve the effectiveness of recommendation systems and leading to the user experience by focusing on model evaluation.

The main problem of CTR prediction is capturing the interactions between user and item features. Existing methods typically address this by employing two approaches such as learning lower-order interactions and deep learning for complex interactions \cite{enhance}. Some methods use Factorization Machines \cite{deepfm},\cite{xdeepfm} to learn second or third order feature interactions in complementary way. And mostly they use deep neural networks (DNNs) based architectures to realize more powerful modeling ability to include high-order feature interactions\cite{dcn}. 

While capturing user-item feature interactions is crucial for CTR prediction, evaluating the effectiveness of these methods are equally important. Here's where the challenge lies: determining the optimal evaluation process to ensure reported performance gains translate to real-world benefits.

The main motivation of this study is to make recommendations both researchers and practitioners for more realistic model evaluations. To do this, we emphasize the importance of model evaluation for recommender systems. Overview the papers on model evaluation, and specifically data split strategies in Introduction. Then, discuss the limitations of the traditional random in Methods section. Continued with the details of the experiments in Experiments section for a better understanding of the experimental results, and also promote reproducible science. Results are given and discussed in Results section. Finally, a summary of the paper is provided in Conclusion.

The contributions of this study are as follows:
\begin{itemize}
\item We compare the performance rankings of 12 state-of-the-art deep-CTR models under random and temporal data split scenarios. No previous studies have examined the performance of the state-of-the-art CTR models using different data split strategies.
\item We demonstrate that model rankings are statistically significantly different under random and temporal data split strategies by conducting a comprehensive statistical analysis.
\item We indicate that there is a temporal data distribution shift in the data by showing the prediction of the test sets which are further away from the training set is harder.
\end{itemize}

\subsection{Model Evaluation}
Evaluating methods to identify the best performers is a crucial phase in building successful recommender systems. This phase assesses a method's effectiveness and its ability to generalize to unseen data. However, determining the optimal evaluation process remains a challenge. With the rapid enhancements in deep-CTR models, it is crucial to assess whether the reported performance gains are genuinely beneficial for the specified tasks. Proper evaluation methodologies are therefore critical to ensure that the tested method will perform well in real-world applications.

Considering the surge in publications proposing advancements in neural network-driven recommendation systems, it is essential for both the researchers and practitioners to prioritize the model evaluation process. This prioritization ensures that any improvements claimed over previous research are genuine. To ensure these advancements in recommender systems translate to real-world benefits, researchers employ both offline and online experimentation methods.

\subsection{Offline and Online Experimentation in Recommender Systems}
Recommender system evaluation involves two main approaches: offline and online experimentation. Online experimentation is not accessible for most researchers. Even when online experimentation is possible, a preliminary assessment is crucial before deploying models for online experimentation to ensure effectiveness. Even a slight misstep in choosing a method for online experimentation can lead to revenue loss and a negative user experience, due to its real-time nature.
Another risk of online experiments is their long feedback loops, meaning it takes time to collect enough data to draw conclusions. Additionally, online experiments require substantial computational resources and can be expensive.

Unlike online experiments, which require real-time user-item interactions, offline experiments are conducted using pre-collected data. As a preliminary step to online testing, offline experiments are crucial for filtering out methods that perform poorly. They serve as a simulation of the online platform \cite{Shani2011}. Offline experimentation allows for rapid iterations of different models and hyper-parameter settings. However, it has limitations, such as its inability to capture the dynamic nature of real-world applications where user preferences constantly change. Therefore, online experiments often follow the successful candidates of offline experiments for CTR prediction before these models receive user traffic.

\section{Previous Work}

\subsection{Model Evaluation in Recommender Systems}

Sun \cite{Sun} explores various data split strategies, arguing that ignoring the dataset's global timeline during evaluation leads to data leakage and an oversimplified representation of user preferences. Castells et al. \cite{Castells} review current offline evaluation protocols in recommender systems to enhance the robustness and resilience of experiments, offering a high-level perspective without delving deeply into specific aspects. Shehzad and Jannach \cite{shehzad2023everyone} critically assess the performance gains reported in state-of-the-art recommender system research, highlighting issues such as the lack of code sharing and inconsistent tuning protocols for baseline and proposed models. Dacrema et al. (2019) \cite{ferrari2019we} question the actual advancements made by recent neural recommendation models, comparing them to simpler methods like top popular recommendations and k-nearest neighbor-based approaches. Dacrema et al. \cite{ferrari2020methodological} examine the methodological flaws in recommender systems research to promote reproducibility and reliability. They emphasize the lack of proper optimization baselines, weak baseline selections, and arbitrary experimental configurations.

\subsection{Data Splitting Strategies in Recommender Systems}

Filipovic et al. \cite{filipovic2020modeling} highlights the significance of incorporating temporal context in user preference modeling to reflect the dynamic nature of real-world recommender systems. However, it advocates for temporal splits without a comprehensive comparison to random splits. Meng et al. \cite{meng2020exploring} compare different data splitting strategies, including random, temporal, and leave-one-out splits, for top-N recommendations. Campos et al. \cite{campos2011towards} stresses the importance of temporal splits for more realistic evaluations in top-N recommendations but focuses narrowly on matrix factorization, a more fundamental approach not directly used in industry.

The aforementioned papers contribute to the recommender systems domain but lack a specific focus on CTR prediction. Our study addresses this gap by concentrating on state-of-the-art CTR prediction models, which are crucial for both academic research and industrial application. Furthermore, we compare model rankings across different experimental settings and investigate model adaptation to temporal data distribution shifts through additional analyses.

While existing studies on model evaluation and data split strategies in recommender systems do not sufficiently address CTR prediction, nor do they compare model rankings or statistically analyze significant differences, our research fills this void. We compare model rankings using a large public dataset, Criteo, and show that evaluations using temporal splits differ significantly from those using random splits. Additionally, we conduct extensive experiments on 12 state-of-the-art deep-CTR models.

Data splitting is a critical step in machine learning, involving the division of available data into training, validation, and test sets. The split strategy significantly impacts model evaluation. Traditional random splits, commonly used in many machine learning fields, leverage randomness in evaluation but fall short with non-stationary data, which depends on time-context.

Temporal splits, dividing data based on time order, offer a more realistic evaluation for non-stationary data. By ensuring models are trained on past data and tested on future samples, temporal splits better reflect real-world applications involving time-dependent data.

\section{Methods}

\subsection{Data Leakage}
One of the key issues with random splits is data leakage. Since the training dataset of a random split may include samples from a later time period than the validation and test datasets, this introduces data leakage into the system. Given that CTR prediction data is non-stationary, randomly splitting the data leads to an inaccurate assessment of the model.

For instance, consider a user who searches for a product, views it on an e-commerce website, adds it to their basket, and finally buys it. This sequence of events reflects the temporal nature of user behavior. Training a user preference predictive model with the user's final purchase action and then asking it to predict if the user would view the product introduces leakage to the system and is unfair. This approach would result in overly optimistic accuracy regarding the user's preferences and lead to not comparing the models accurately.

\begin{figure}[h]
    \centering
    \includegraphics[width=0.5\linewidth]{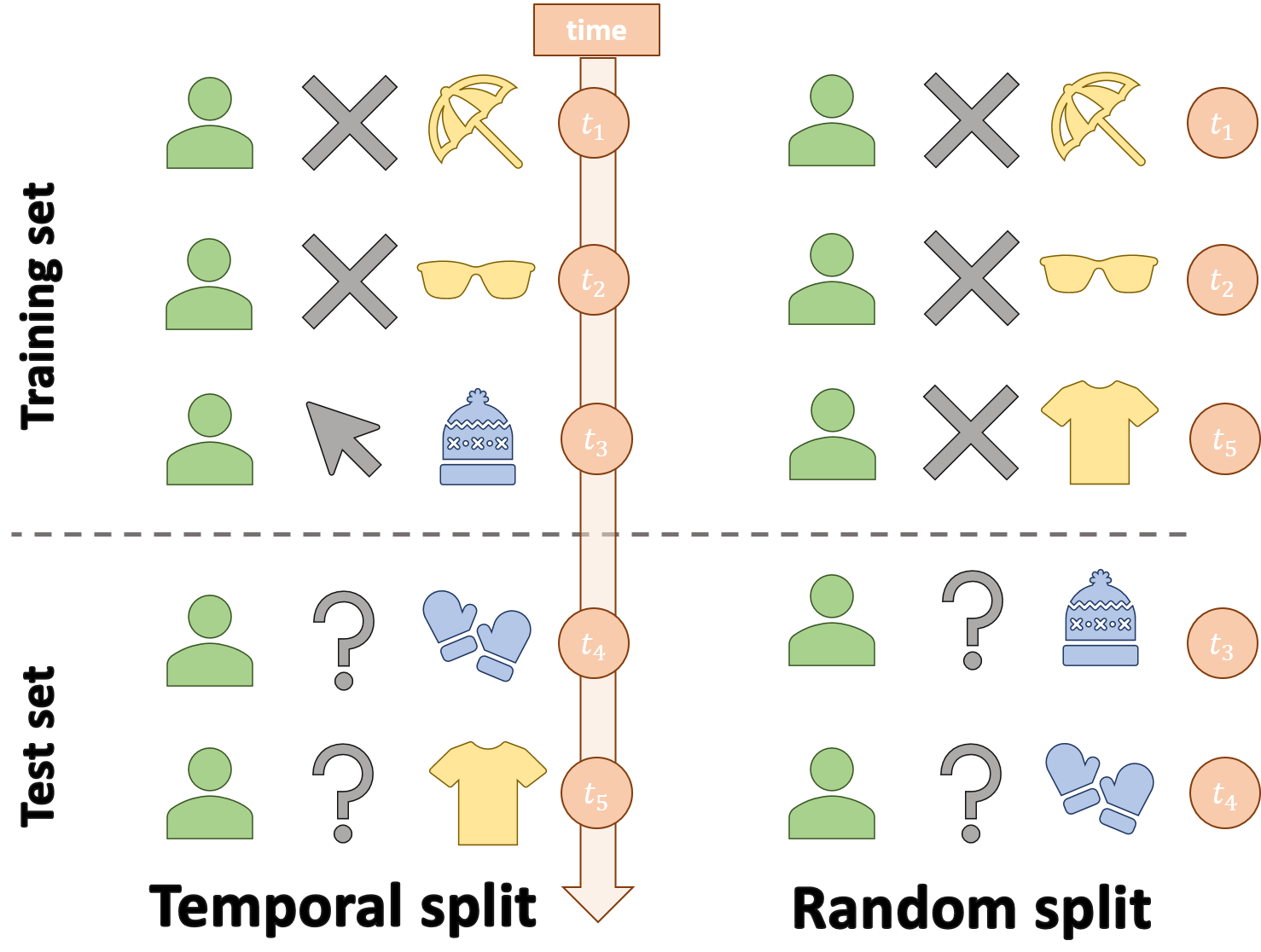}
    \caption{An illustration of temporal patterns of user-item interactions. The interaction at time $t_{5}$ leaks to test set in random split. Moreover, model can learn from temporal patterns in temporal split, such as the user clicked to winter clothes while not clicking to summer clothes.}
    \label{fig:figure_1}
\end{figure}

\subsection{Capturing Temporal Patterns}
In contrast to random splits, temporal splits prevent data leakage by directly simulating real-world applications in terms of time. By preserving the temporal order, temporal splits allow models to capture temporal dynamics. This approach ensures that models capable of understanding such patterns perform well in offline experiments, leading to more indicative results for algorithm selection.

User preferences evolve over time, making the data non-stationary and filled with temporal patterns. Random splits fail to account for these temporal dynamics. Another aspect is that since random splits do not consider the chronological order of the samples, they cannot effectively measure a model's ability to capture the temporal dynamics inherent in the data.

Remember the example of the user that searches for ad product and as a result of the sequence of the events, buys the product. Conversely to random split, if the model is fed with the user's sequential actions, it will enable the model to learn from these sequential actions, and test the model's ability to capture the relationship between viewing, adding to the basket, and purchasing behaviors.

\subsection{Concept Drift}
To train recommender systems, practitioners typically collect long-term data to ensure that user-item interactions are generalizable. However, data distribution changes over time, a phenomenon known as concept drift, which significantly affects model performance \cite{Gama, schlimmer1986beyond}. Models can adapt to concept drift by understanding temporal patterns. Various studies have explored methods to overcome concept drift. Turksoy et al. \cite{TIF} introduce a loss function that penalizes the mispredictions of recent samples more heavily by weighting the losses using different mathematical variants. Li et al. \cite{Li_concept} utilize a time-interval attention layer to account for the drift in item popularity and user preferences over time. Koychev \cite{Koychev} introduces sample weights based on viewing frequency at a given time and implemented a forgetting mechanism for older samples.

\section{Experiments}

\subsection{Dataset}
Criteo dataset is a widely-used benchmark CTR prediction dataset, containing one week of display advertising data. It includes 45,840,617 user-item interactions with 13 numerical and 26 categorical features. We followed the data preprocessing steps of \cite{afn}. The dataset is divided into training, validation, and test sets as 7:2:1, respectively.

\begin{figure}[h]
    \centering
    \includegraphics[width=0.8\linewidth]{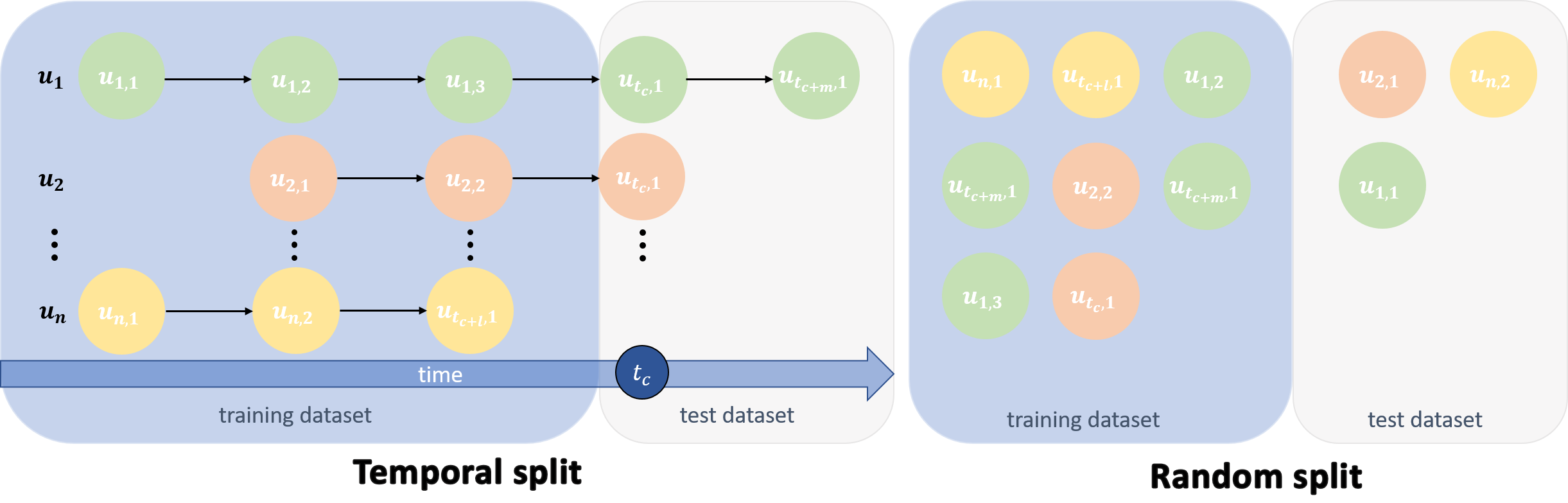}
    \caption{Illustrations of temporal split, and random split.}
    \label{fig:figure_2}
\end{figure}

For reproducibility, we utilized the dataset provided by the BARS benchmark for random split
\footnote{\url{https://huggingface.co/datasets/reczoo/Criteo_x1/tree/main}.}.
For temporal split, we used the original version released by Criteo, which is chronologically ordered. The split ratio and preprocessing steps for the temporal split are consistent with those used for the random split
\footnote{\url{https://go.criteo.net/criteo-research-kaggle-display-advertising-challenge-dataset.tar.gz}.} The representation of these splits is provided in Figure \ref{fig:figure_1}.

\subsection{Models}

\begin{itemize}
    \item FinalMLP \cite{finalmlp} is an effective model using two well-tuned MLPs, gets further boosted with layers improving how it handles features and combines interactions.
    \item DNN \cite{dnn} has single multilayer perceptron to capture high-order interactions.
    \item MaskNet \cite{masknet} introduces a specific multiplication operation  with instance-guided masks to capture feature interactions.
    \item AFN+ \cite{afn} learns complex feature interactions adaptively from data to handle noisy feature combinations.
    \item DCNv2\cite{dcnv2} uses cross network in learning bounded-degree feature interactions while keeping the benefits of a DNN model.
    \item IPNN \cite{ipnn} utilizes product layer to capture how features from different categories interact.
    \item DeepIM \cite{deepim} combines interaction machine, capturing high order interactions in a simple and fast way, with deep neural networks.
    \item EDCN \cite{edcn} a model with lightweight information-sharing modules to share information effectively between separate networks.
    \item AOANet \cite{aoanet} proposes a paradigm that allows the model to learn and optimize the interaction operation itself based on the data, essentially searching for the most effective way to combine features for a specific task.
    \item xDeepFM \cite{xdeepfm} introduces Compressed Interaction Network (CIN) which captures feature interactions explicitly and efficiently at the vector level.
    \item Fibinet \cite{fibinet} dynamically prioritizes features using a Squeeze-Excitation Network (SENET) technique and captures fine-grained interactions between features with bilinear functions.
    \item Wide\&Deep \cite{widedeep} leverages a wide linear model to explicitly capture specific feature interactions.
\end{itemize}

\subsection{Implementation}
FuxiCTR version 2.0.1 with PyTorch 2.3.0 is used for the offline experiments on Criteo dataset. Software details of other libraries used for the implementation are given in Appendix. to promote reproducible science. \par
We repeate the experiments three times, and present the mean and standard deviation values in Results section. For reproducilibity, 2021, 2022, and 2023 are used as random seed in the model configurations of FuxiCTR.

\subsection{Evaluation Metrics and Statistical Analysis}
To assess the models, we utilize the most commonly used performance metrics for CTR prediction: AUC (Area Under the Curve) and logloss. Additionally, a paired t-test is conducted to compare the performance metrics under different splits. Spearman's correlation coefficient and Kendall's $\tau$ are employed to test the hypothesis that the rankings of the deep-CTR models differ between random and temporal data split strategies.

\section{Results}

\begin{table}[h]
\centering
\caption{Experimental results and ranking changes of 12 state-of-the-art deep-CTR models under random, and temporal splits.}
\label{tab:table_1}
\resizebox{\textwidth}{!}{%
\begin{tabular}{cccc|cccc}
\hline
\multicolumn{4}{c|}{{\color[HTML]{000000} \textbf{random split}}} & \multicolumn{4}{c}{{\color[HTML]{000000} \textbf{temporal split}}} \\ \hline
\textbf{ranking}  & \textbf{model}  & \textbf{AUC}  & \textbf{logloss}  & \textbf{ranking}   & \textbf{model}   & \textbf{AUC}  & \textbf{logloss}  \\
1  & FinalMLP   & 0.81486 ± 0.00007 & 0.43715 ± 0.00007 & 1 ($\leftrightarrow$)                               & FinalMLP   & 0.80604 ± 0.00042 & 0.44829 ± 0.00092 \\
2  & EDCN       & 0.81432 ± 0.00012 & 0.43763 ± 0.00009 & 2 ($\uparrow$ 7)    & DNN        & 0.80478 ± 0.00099 & 0.44937 ± 0.00097 \\
3  & AFN+       & 0.81411 ± 0.00011 & 0.43780 ± 0.00012 & 3 ($\uparrow$ 3)    & MaskNet    & 0.80466 ± 0.00039 & 0.44895 ± 0.00039 \\
4  & AOANet     & 0.81408 ± 0.00002 & 0.43783 ± 0.00005 & 4 ($\downarrow$ 1)  & AFN+       & 0.80374 ± 0.00060 & 0.44968 ± 0.00051 \\
5  & DeepIM     & 0.81391 ± 0.00005 & 0.43798 ± 0.00004 & 5 ($\uparrow$ 7)    & DCNv2      & 0.80309 ± 0.00012 & 0.45116 ± 0.00059 \\
6  & MaskNet    & 0.81373 ± 0.00011 & 0.43818 ± 0.00013 & 6 ($\uparrow$ 4)    & IPNN       & 0.80091 ± 0.00317 & 0.45286 ± 0.00302 \\
7  & Wide\&Deep & 0.81373 ± 0.00010 & 0.43812 ± 0.00010 & 7 ($\downarrow$ 1)  & DeepIM     & 0.79821 ± 0.00373 & 0.45648 ± 0.00392 \\
8  & xDeepFM    & 0.81368 ± 0.00010 & 0.43816 ± 0.00012 & 8 ($\downarrow$ 6)  & EDCN       & 0.79927 ± 0.00239 & 0.45644 ± 0.00186 \\
9  & DNN        & 0.81360 ± 0.00008 & 0.43823 ± 0.00005 & 9 ($\downarrow$ 5)  & AOANet     & 0.79814 ± 0.00128 & 0.45512 ± 0.00117 \\
10 & IPNN       & 0.81352 ± 0.00014 & 0.43840 ± 0.00015 & 10 ($\downarrow$ 2) & xDeepFM    & 0.78559 ± 0.01138 & 0.48638 ± 0.04263 \\
11 & FiBiNET    & 0.81298 ± 0.00015 & 0.43887 ± 0.00020 & 11 ($\leftrightarrow$)                             & FiBiNET    & 0.77636 ± 0.01859 & 0.49393 ± 0.04850 \\
12 & DCNv2      & 0.81367 ± 0.00060 & 0.43837 ± 0.00060 & 12 ($\downarrow$ 5) & Wide\&Deep & 0.78023 ± 0.01071 & 0.49077 ± 0.03989 \\ \hline
\end{tabular}%
}
\end{table}

The results of the comprehensive experiments of 12 state-of-the-art models are presented in Table \ref{tab:table_1}. The results show that 

Paired t-test on AUCs of the models on different splits show that model performances are statistically significantly different in random and temporal splits ($p<0.0001$)

Spearman's correlation coefficient of 0.2378 and Kendall's $\tau$ correlation of -0.0606 indicate that there is a weak correlation between the model rankings in random and temporal splits with many rank swaps.

\begin{table}[h]
\centering
\caption{Comparative study of different test datasets split by temporal order. no.1, no.2, no.3, and no.4 test splits are respectively the 7th, 8th, 9th, 10th parts of the whole dataset.}
\label{tab:table_2}
\begin{tabular}{lccc}
\hline
\multicolumn{1}{c}{\textbf{DNN}}      & \textbf{AUC} & \textbf{logloss} & \textbf{\(\Delta \)AUC in consecutive days} \\ \hline
split \#1 & 0.80107 ± 0.79500 & 0.45268 ± 0.45841 & \textbf{} \\
split \#2 & 0.80185 ± 0.80606 & 0.45477 ± 0.45058 & -0.36\%   \\
split \#3 & 0.79872 ± 0.80318 & 0.45758 ± 0.45325 & -0.57\%   \\
split \#4 & 0.79690 ± 0.80147 & 0.45255 ± 0.44826 & -0.83\%   \\ \hline
\multicolumn{1}{c}{\textbf{FinalMLP}} & \textbf{AUC} & \textbf{logloss} & \textbf{\(\Delta \)AUC in consecutive days} \\ \hline
split \#1 & 0.80606 ± 0.00368 & 0.45058 ± 0.00368 &           \\
split \#2 & 0.80318 ± 0.00390 & 0.45325 ± 0.00379 & -0.36\%   \\
split \#3 & 0.80147 ± 0.00399 & 0.44826 ± 0.00376 & -0.57\%   \\
split \#4 & 0.79936 ± 0.00381 & 0.45417 ± 0.00372 & -0.83\%   \\ \hline
\end{tabular}
\end{table}

To further explore concept drift and the models' ability to adapt to recent trend changes over time, an analysis is conducted. The entire dataset is divided into 10 equal-sized parts. The training set comprises the first 50\% of the dataset (the first 5 parts). The validation set consists of the next 10\% (the 6th part). The test sets are the subsequent 10\% segments, specifically the 7th, 8th, 9th, and 10th parts of the dataset. These test splits are named as split \#1, split \#2, split \#3, and split \#4, respectively. FinalMLP and DNN models, which are the best two models in temporal split, are used for this purpose.

The results are presented in Table \ref{tab:table_2}. It is obvious that as the test dataset moves away from the training set in time axis, the AUCs decrease for both DNN and FinalMLP models.

\section{Conclusion}
In this paper, we present a comparative study of different data splitting strategies: random and temporal. We discuss the disadvantages of random splits, such as data leakage and inadequate assessment of a model's ability to capture temporal dynamics, in the context of real-world applications. \par
Our experimental results demonstrate that the overall outcomes differ statistically significantly between random and temporal splits. The higher AUC values observed in the random split suggest the presence of data leakage. Detailed statistical analysis indicates significant differences in model rankings depending on the data split strategy used. Thus, considering the real-world application of CTR prediction, we recommend that researchers and practitioners incorporate temporal splits in their model evaluation processes, as it better reflects real-world conditions. \par
We have examined the effects of data splitting strategies in offline experiments. Given that CTR prediction is an industrial task, future work could enrich these findings with online experiments using company dataset.

\newpage
\bibliographystyle{unsrt}
\bibliography{references}

\section{Appendix}

\subsection{Software Details}
\begin{verbatim*}
cuda==12.3
python==3.9.13
pytorch==2.3.0
pandas==1.4.4
numpy==1.24.4
scipy==1.13.1
sklearn==1.0.2
pyyaml==6.0
h5py==3.7.0
tqdm==4.64.1
fuxictr==2.0.1
\end{verbatim*}
\end{document}